\documentclass[aps,prl,twocolumn]{revtex4}
\usepackage{graphicx}
\usepackage{amsmath}
\usepackage{mathtools}
\usepackage{slashed}
\newcommand{\BlackHat}{{\sc BlackHat}}
\newcommand{\SHERPA}{{\sc SHERPA}}

\newcommand{\COMIX}{{\sc COMIX}}

\newcommand{\Maitre}{Ma\^{\i}tre}

\newif\ifdraft
\drafttrue
\newif\ifpreprint
\preprintfalse

\def\Fig#1{Fig.~{\ref{#1}}}

\def\tab#1{table~{\ref{#1}}}

\def\bom#1{{\mbox{\boldmath $#1$}}}
\def\WW{$WW$}

\def\WWjn{$WW\,\!+\,n$}

\def\WpWm{$W^+W^-$}

\def\WpWmjjj{$W^+W^-\!+3$}
\def\WpWmjjjj{$W^+W^-\!+4$}

\def\WpWmjjx{$W^+W^-\!+1,2$}
\def\WpWmjn{$W^+W^-\!+n$}
\def\WpWmjnm{$W^+W^-\!+(n\!-\!1)$}

\def\Wjnp1{$W\,\!+\,(n\!+\!1)$}

\def\jet{{\rm jet}}
\def\pt{p_T}

\def\HTpartonic{{\hat H}_T}

\def\LCAccTot{1\%}   

\def\ScaleLOjjjj{50\%}
\def\ScaleLOjjj{40\%}

\def\ScaleLO{7\%}
\def\ScaleNLOjjj{5\%}

\newbox\charbox
\newbox\slabox
\def\s#1{{      
        \setbox\charbox=\hbox{$#1$}
        \setbox\slabox=\hbox{$/$}
        \dimen\charbox=\ht\slabox
        \advance\dimen\charbox by -\dp\slabox
        \advance\dimen\charbox by -\ht\charbox
        \advance\dimen\charbox by \dp\charbox
        \divide\dimen\charbox by 2
        \raise-\dimen\charbox\hbox to \wd\charbox{\hss/\hss}
        \llap{$#1$}
}}

\begin{document}

\title{
\ifpreprint
\hbox{\rm\small
FR-PHENO-2015-005
\break}
\hbox{$\null$\break}
\fi
$\bom{W^+W^-}$\,+\,3 Jet Production at the Large Hadron Collider in NLO QCD
}

\author{
	F.~Febres Cordero,
        P. Hofmann and
        H.~Ita 
\\
$\null$
\\
Physikalisches Institut, Albert-Ludwigs-Universit\"at Freiburg\\
       D--79104 Freiburg, Germany
}

\begin{abstract}
We present next-to-leading order (NLO) QCD predictions to \WpWm{} production in
association with up to three jets at hadron colliders. We include contributions
from couplings of the $W$ bosons to light quarks as well as trilinear vector
couplings.  These processes are used in vector-boson coupling measurements,
are background to Higgs signals and are needed to constrain many new physics
scenarios.
For the first time NLO QCD predictions are shown for electroweak di-vector boson production
with three jets at a hadron collider. 
We show total and differential cross sections for the LHC with proton
center-of-mass energies of 8 and 13~TeV.
To perform the calculation we employ on-shell and unitarity methods implemented 
in the \BlackHat{} library along with the
\SHERPA{} package. We have produced event files that can be accessed for future
dedicated studies. 
\end{abstract}

\maketitle

The increasing energy reach of the Large Hadron Collider (LHC) and the large
datasets expected from the ATLAS and CMS experiments widen the need for precise
theoretical predictions. 
Scattering processes including final-state leptons, produced through
intermediate electroweak vector bosons ($W^\pm$, $Z$ or $\gamma$), 
with multiple jets are particularly important as they probe the
electroweak sector of the Standard Model (SM) and are central to many
searches for new particles. 
Precise theory predictions are necessary in order to challenge 
the SM and also to extend the experiments' reach by providing a reference
for backgrounds to rare events. 
Here we study high-multiplicity signatures
which open new perspectives on fundamental interactions. 
They can arise when several particles recoil against the core
scattering processes, exposing the dynamics of fundamental
interactions. Furthermore, such signatures can be the outcome of decay chains of
potential new heavy particles.  
The comparison of precise theoretical predictions to experimental measurements will open
numerous ways to sharpen our understanding of physics at the TeV scale.

The production of two oppositely charged $W$ bosons in association to several
jets stands out due to its rich phenomenology. This final-state
configuration is obtained in top-quark pair production, as the top quarks decay
into bottom quarks and $W$ bosons. 
It also appears in the weak vector-boson-fusion (VBF) mechanism, in which
scattering quarks emit intermediate vector bosons which scatter and
eventually produce a \WpWm{} pair. 
The VBF production mode of the Higgs boson which decays to $W$ bosons
contributed to early limits of the Higgs-boson mass~\cite{VBFHWWmass} and 
helped the measurement of its couplings and spin~\cite{VBFHWWspincouplings}.
The distinguishing signature of these VBF processes are two forward pointing
jets (arising from the emitter quarks) including a radiation gap between the
jets.  Precise predictions for \WpWmjjj{}-jet production allow to explore
this gap in detail.
Moreover, predictions for \WpWmjn{}-jet processes are important for
measuring trilinear and quartic vector-boson couplings.  
Finally, in many scenarios of physics beyond the SM these leptonic states appear
in combination with many jets as end products of decay chains of heavy colored
particles.

In this Letter, we present NLO QCD predictions for \WpWmjjj{}-jet production,
which is the first time this level of precision is achieved for processes of
electroweak di-vector boson production in association with three jets. Results
for \WpWmjn-jet production with $n=0,1$ and $2$ are shown as well. 
We include contributions from couplings of the vector bosons to light
quarks (omitting top contributions) as well as trilinear vector couplings. We
consider the leading electroweak contributions.
A phenomenological study is presented to assess the impact of NLO QCD
corrections on the total and differential cross sections at the LHC with collision
energies of $8$ and $13$ TeV. 
The reliability of the theory predictions is significantly improved by including QCD corrections,
as uncertainties for \WpWmjjj{}-jet total cross sections are reduced from \ScaleLOjjj{} at
leading order (LO) to about \ScaleNLOjjj{} at NLO.

Measurements of the \WWjn{}-jet production cross section ($n\ge 0$) have a long
history at hadron colliders.
The inclusive \WW{} cross section was measured first by the
CDF~\cite{WWCDF} and D0~\cite{WWD0} experiments at the Tevatron, and recently by
the ATLAS~\cite{WWATLAS,WWATLASPRD} and
CMS~\cite{WWCMS7TeV,WWCMS8TeV} experiments at the LHC.
Very recently a first dedicated measurement of the cross sections for \WpWm{}
production in association with $0$, $1$ and $2$ jets has been
published~\cite{WWjCDF} by the CDF experiment.
In the context of the exploration of the electroweak sector, many studies have
been carried out to measure (and in cases subtract) direct \WWjn{}-jet signals.
For example, \WpWmjn{}-jet production has been thoroughly studied in order to
reduce its contributions (through kinematic constrains and jet vetoes) as
background for Higgs-boson studies in the \WpWm{} decay
mode~\cite{HWWCDF,HWWD0,HWWCMSJHEP,HWWATLASPRD}.
A year ago both ATLAS~\cite{WWjjATLAS} and CMS~\cite{WWjjCMS} showed first
evidence for the production of same charge vector-boson scattering.
In the future dedicated studies at the LHC of \WpWm{}+jets production in bins
of jet multiplicity should be carried out, in order to assist measurements of Higgs-boson
properties and couplings of massive vector bosons.  It is the aim of this work
to provide reliable SM predictions for such a task.

The importance of precise theoretical predictions for inclusive \WpWm{} pair production
has stimulated many developments starting from the early LO
results~\cite{WWLO}, including NLO QCD corrections~\cite{WWNLO} and, very
recently, the first next-to-next-to-leading (NNLO) order QCD prediction
\cite{WWNNLO}. For \WpWm{} pair production in association with jets, NLO QCD
predictions are available for one~\cite{WW1j} and two
jets~\cite{WW2ja,WW2jb,Alwall:2014hca} (including same sign studies
in ref.~\cite{WpWp2j}).  NLO QCD corrections are particularly large at low jet
multiplicity due to the opening of additional partonic channels and the release of kinematic constrains. In general, the
QCD corrections significantly reduce the unphysical factorization- and
renormalization-scale dependence, and provide then the first quantitatively
reliable predictions for total cross sections and shapes of distributions. 

The virtual corrections to \WpWmjjj{}-jet production
involve up to five final-state
objects. Counting the leptonic decay products of the
weak vector bosons, this process contains a phase space with up to seven
partonic/leptonic final states. This is a state-of-the-art NLO QCD calculation
for processes with five or more objects in the final
state~\cite{W5jBH,W4jBH,Z4jBH,OtherNLO2to5}.
We employ the \BlackHat{} library~\cite{BlackHatI} to compute the one-loop matrix
elements. This library has been developed based on on-shell
methods~\cite{UnitarityMethod,BCFUnitarity,BCFW,OPPEtc,OnShellReview}
that rely on the unitarity and factorization properties of scattering
amplitudes to construct loop and tree amplitudes. Such methods are efficient 
and scale well as the number of external legs
increases. In particular the \BlackHat{} library has produced
NLO QCD predictions to vector-boson production in association with up to five
jets \cite{W3jBH,W3jDistributions,W4jBH,W5jBH,Z3jBH,Z4jBH,PhotonBH}, pure
$4$-jet production~\cite{FourJetsBH} and di-photon production in association
with $2$ jets~\cite{yyBH}.

\begin{figure}[t]
\centering
\includegraphics[clip,scale=0.7]{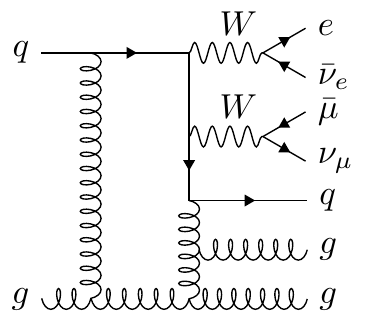}
\includegraphics[clip,scale=0.7]{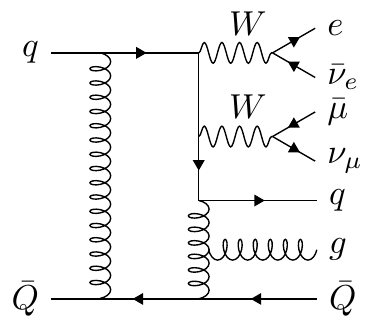}
\includegraphics[clip,scale=0.7]{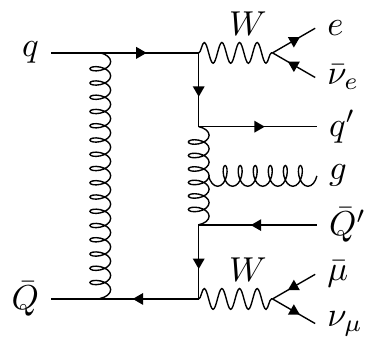}
\newline
\includegraphics[clip,scale=0.8]{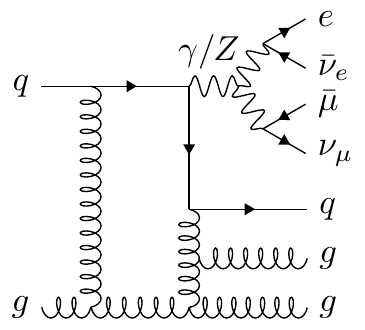}
\includegraphics[clip,scale=0.8]{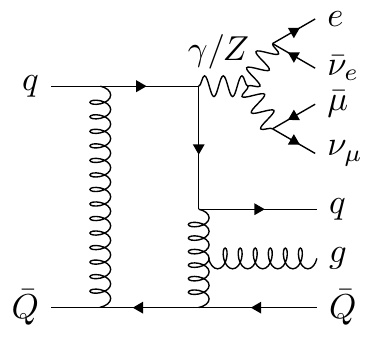}
\caption{Sample diagrams of loop amplitudes for
$q g \rightarrow W^+W^- \, q^{(\prime)} \! g g$ and 
$q \bar Q \rightarrow W^+W^- \, q^{(\prime)} \! g \bar Q^{(\prime)}$, followed by
$W^- \to e\bar\nu_e$ and $W^+\to \bar\mu\nu_\mu$.} 
\label{lcdiagramsFigure}
\end{figure}
%

In this work we extended the functionality of the
\BlackHat{} library to provide virtual matrix elements for QCD corrections to
two massive vector bosons and up to five partons (gluons or quarks). To this
end on-shell recursion relations~\cite{BCFW} for tree amplitudes of quarks,
gluons and massive vector bosons~\cite{VbosonTrees} (including decay products)
have been provided.  As a cross check, we implemented the same tree amplitudes
through off-shell recursions~\cite{BGrec}. Both implementations have been
compared yielding numerically identical results. In addition, we provided
infrastructure to compute loop amplitudes based on the new tree input.
Finally, tree and loop amplitudes are assembled in an automated way~\cite{W4fc}
into the virtual, squared matrix elements. 

\begin{table*}
\vskip .4 cm
\begin{tabular}{||c||c|c||c|c||c|c||c|c||}
\cline{2-9}
\multicolumn{1}{c|}{} &  \multicolumn{2}{c||}{\WpWmjn{} jet (8 TeV)} & \multicolumn{2}{c||}{$R_n$ (8 TeV)} & \multicolumn{2}{c||}{\WpWmjn{} jet (13 TeV)} & \multicolumn{2}{c||}{$R_n$ (13 TeV)} \\  \hline
$n$ &  LO & NLO & LO & NLO  &  LO & NLO & LO & NLO \\  \hline\hline
0 & $142.2(3)^{+3.7}_{-5.3}$  & $207.4(7)^{+5.1}_{-3.3}$  & --- & --- & $230.7(5)^{+13.7}_{-16.7}$  & $358(2)^{+7.3}_{-4.5}$ & --- & ---  \\  \hline
1 & $60.9(1)^{+9.8}_{-8.0}$   & $76.0(2)^{+3.6}_{-3.9}$  & $0.428(1)$ & $0.366(2)$ & $131.6(2)^{+16.3}_{-14.0}$  & $165.1(6)^{+7.2}_{-7.1}$ & $0.571(2)$ & $0.462(3)$  \\  \hline
2 & $29.43(6)^{+9.99}_{-6.91}$   & $28.5(1)^{+0.4}_{-1.8}$  & $0.483(1)$ & $0.376(2)$ & $77.5(2)^{+23.1}_{-16.6}$   & $72.7(4)^{+0.2}_{-3.2}$  & $0.589(2)$ & $0.440(3)$ \\  \hline
3 & $11.11(2)^{+5.73}_{-3.51}$   & $9.05(12)^{+0.08}_{-0.90}$   & $0.378(1)$ & $0.317(5)$ & $35.59(6)^{+16.66}_{-10.55}$   & $28.1(3)^{+0.0}_{-2.1}$ & $0.459(1)$ & $0.386(5)$ \\  \hline
4 &  $3.58(1)^{+2.49}_{-1.37}$   & ---  & $0.322(1)$  & --- &  $14.12(8)^{+9.05}_{-5.14}$   & ---  & $0.397(2)$  & --- \\  \hline

\end{tabular}
\caption{Total cross sections in femtobarns (fb) for \WpWmjn{}-jet production at
the LHC with $\sqrt{s}=8$ and 13~TeV. We show parton level LO and NLO QCD results, as
well as the corresponding jet cross section ratios, $R_n=$(\WpWmjn{} jet) / (\WpWmjnm{} jet). The NLO result for
\WpWmjjj{} jets uses the leading-color approximation discussed in the text.  In
parentheses we show the numerical integration error, and the dependence on the
unphysical renormalization and factorization scales is quoted in super and
subscripts.
\label{CrossSectionLHC} 
}
\end{table*}

We show representative virtual diagrams in \Fig{lcdiagramsFigure}.  
We use a
leading-color approximation in the finite virtual contributions~\cite{W4fc},
including virtual quark loops, but dropping other sub-leading color terms in the
partial loop amplitudes. 
We keep the full
color dependence in born and real radiation contributions.  We have confirmed that this
approximation is an excellent one for \WpWmjjx-jet production, shifting the
total cross section below \LCAccTot, 
less than uncertainties from parton
distributions or higher-order terms in $\alpha_s$.  
The omitted contributions correspond to $N_c$-suppressed gluon-loop diagrams,
diagrams with vector bosons coupled directly to closed quark loops, 
as well as diagrams with a massive top quark in the loop.
In fact, the latter two contributions have been shown to shift the total
cross section below the percent-level~\cite{WW2jb}. 
We therefore expect the omitted sub-leading color corrections to 
\WpWmjjj-jet production to be small. 

We consider the double resonant contributions and include the full Breit-Wigner
distribution for intermediate $Z$ and $W$ bosons; decays to leptons retain all
spin correlations.  We consider a diagonal Cabibbo-Kobayashi-Maskawa (CKM)
matrix and for loop diagrams we consider five light quark flavors;
$n_f=5$. Except for canceling infrared collinear poles, we do not include 
contribution from external bottom quarks; bottom final states are vetoed being
naturally attributed to studies of top production which we do not consider
here. Given the sub-leading nature of bottom parton-distribution functions
(PDFs), di-bottom initial states are small contributing at the 
percent
level~\cite{WW1j} at low multiplicities justifying the approach.

The remaining NLO ingredients, the real-emission and dipole-subtraction
terms~\cite{CS}, are computed by \COMIX{}~\cite{Comix}, part of the \SHERPA{}
package~\cite{Sherpa}.  We also use \SHERPA{} to perform phase-space
integration. We store intermediate results in public root-format~\cite{ROOT}
ntuple files~\cite{BHntuples}, recording parton momenta, along with
the information on scale-, coupling- and parton-distribution-function dependent
contributions to the event weights.  The effects of varying scale and input
parameters as well as tightening of phase-space cuts are obtained efficiently
using the ntuple files without recomputing the time-consuming matrix elements.

%

The results have been checked by observing the explicit cancellation of 
infrared poles,
comparison of one-loop matrix elements at individual phase-space points where
available~\cite{Cullen:2014yla}, as well as comparison to an implementation 
using off-shell
recursion relations~\cite{BGrec}. The \SHERPA{} implementation of the
real-emission and dipole-subtraction terms has a dependence on an arbitrary parameter
$\alpha_{\rm dipole}$ \cite{AlphaDipole}, and we
have made dedicated studies to show that our results are independent of it.

\begin{figure*}[t]
\includegraphics[clip,scale=0.57]{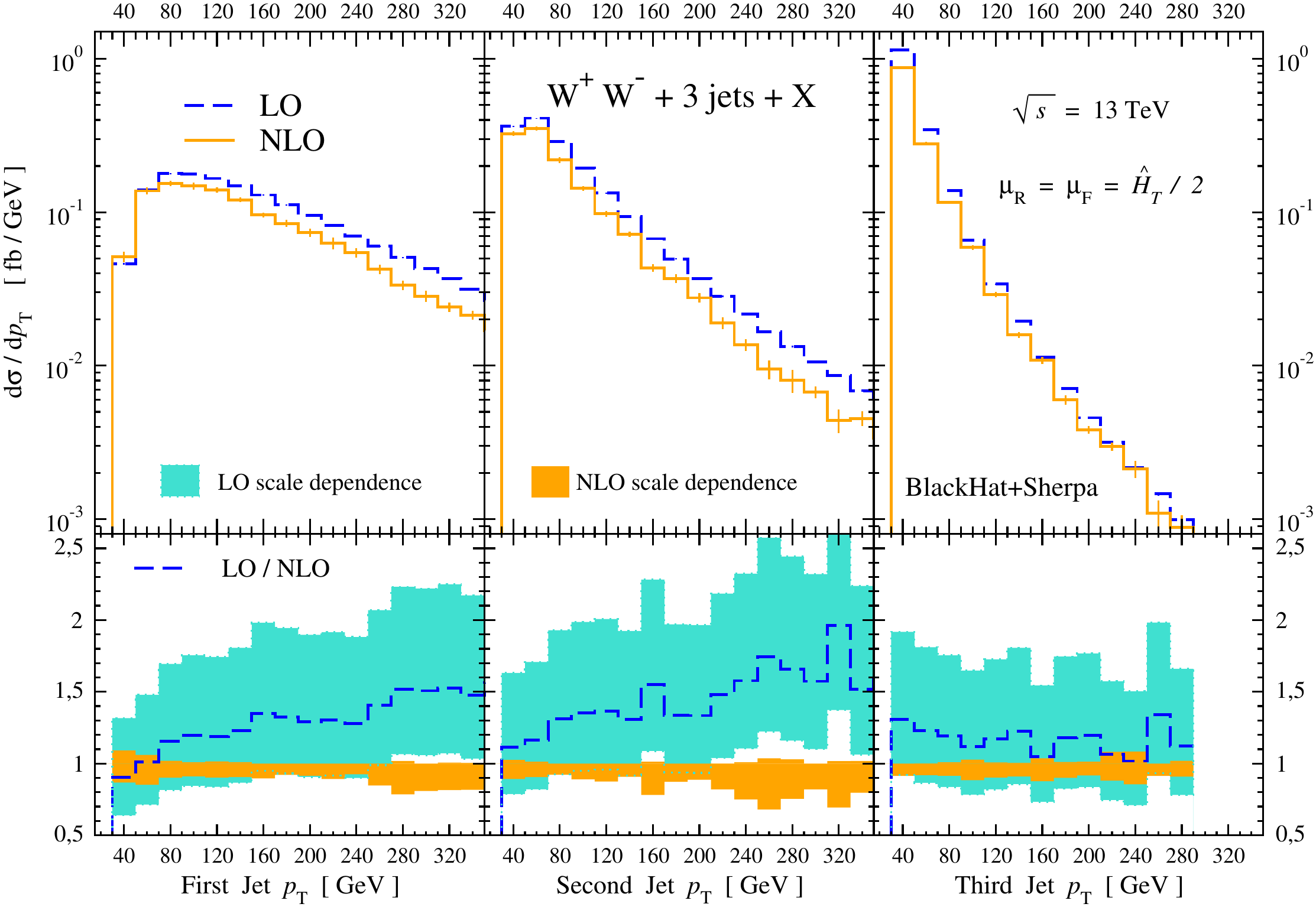}
\caption{A comparison of the $p_T$ distributions of the leading three
jets in \WpWmjjj{}-jet production at the LHC at $\sqrt{s}=13$~TeV.  In
the upper panels the NLO distribution is the solid (orange) histogram
and the LO predictions are shown as dashed (blue) lines.  The thin
vertical line in the center of each bin (where visible) gives its
numerical (Monte Carlo) integration error.  The lower panels show the
LO distribution and LO and NLO scale-dependence bands 
normalized to the central NLO prediction.
The bands are shaded (orange) for NLO and light-shaded
(cyan) for LO. 
}
\label{WW3ptFigure}
\end{figure*}

In our study, we consider the inclusive processes $p p \rightarrow$ \WpWmjn{}
jets ($n=0,1,2$ and $3$, and also $n=4$ at LO) at the LHC with proton center-of-mass energy
of $\sqrt{s} = 8$ and 13~TeV. 
The $W$ bosons are decayed into oppositely charged
light lepton-neutrino pairs; we require an electron as well as an anti-muon.
We consider leptons from distinct flavors to distinguish from processes with pairs
of $Z$ bosons, $pp\rightarrow ZZ (\rightarrow\ell^+\ell^-\nu\bar\nu)+$jets.
We set dynamically the renormalization and factorization scales according to 
$\mu = \mu_r=\mu_f= \HTpartonic/2$, where $\HTpartonic = \sum_j p_T^j$ and the sum
runs over all final-state partons and leptons labelled by $j$.
Scale-dependence bands are constructed from the
minimum and maximum of the given observable evaluated at five values: $\mu/2,
\mu/\sqrt2, \mu, \sqrt2\mu, 2\mu$.
The following
phase space cuts are applied: $p_{\rm T}^{e,\mu} > 20$~GeV, $|\eta^{e,\mu}| < 2.4$,
$\slashed{E}_{\rm T}>30$~GeV, $p_{\rm T}^{e\mu}>30$~GeV and $m_{e\mu}>10$~GeV.
Jets are defined with the anti-$k_{\rm T}$ jet algorithm~\cite{antikT} with
$R=0.4$ and requiring $\pt^\jet > 30$~GeV, $|\eta^\jet|<4.5$, and are ordered
in $\pt$. 
Here, $p_T$ are transverse momenta; $\eta$, pseudo rapidities; $m_{e\mu}$, and
$p_T^{e\mu}$ the mass and transverse momentum of the electron-muon system,
respectively.  We identify $\slashed{E}_{\rm T}$ with the magnitude of the sum
of the momenta of the neutrinos in the transverse plane. 
At NLO we use the
MSTW2008nlo~\cite{MSTW2008} PDFs and the MSTW2008lo set
at LO, and we set $\alpha_s$ consistently as provided by the PDFs.  
The $W$ and $Z$ boson mass
and width are given respectively by $\Gamma_W=2.085$~GeV, $M_W=80.399$~GeV and
$\Gamma_Z=2.4952$~GeV, $M_Z=91.188$~GeV. 
We use the complex-mass scheme~\cite{ComplexMS} in order to treat the unstable vector
bosons in a gauge invariant way.

In \tab{CrossSectionLHC}, we present LO and NLO parton-level cross sections
at $\sqrt{s}=8$ and 13~TeV for inclusive \WpWm{} production accompanied by zero through three jets.
In addition we show LO results for \WpWmjjjj{} jets.
The scale-variation for the NLO
cross sections amount to $\ScaleNLOjjj$ for our setup
as opposed to the LO values which vary from $\ScaleLO$ to $\ScaleLOjjjj$  
for zero up to four jets at $13$~TeV. 
We also display 
the ``jet-production'' ratios of
\WpWmjn-jet to \WpWmjnm-jet production. 
These kinds of ratios are less sensitive to experimental and theoretical
systematics than the absolute cross sections.  Despite a significant dependence
of the ratios on the phase-space cuts imposed~\cite{Wratios} universal behavior
of these ratios is expected for large jet multiplicities. 
More detailed studies of these ratios for
higher jet multiplicities at NLO shall be performed.

As expected total cross sections increase with the center-of-mass energy (see
\tab{CrossSectionLHC}).  As phase-space cuts are maintained the
increase in center-of-mass energy opens more phase-space volume and enlarges
the range of momentum fractions of partons sampled in the incoming protons. 
We observe that the jet ratios of $W^+W^-+n$-jet production increase with
the collision energy consistent with the fact that additional jets have 
more phase space available.

\begin{figure}[t]
\includegraphics[clip,scale=0.37]{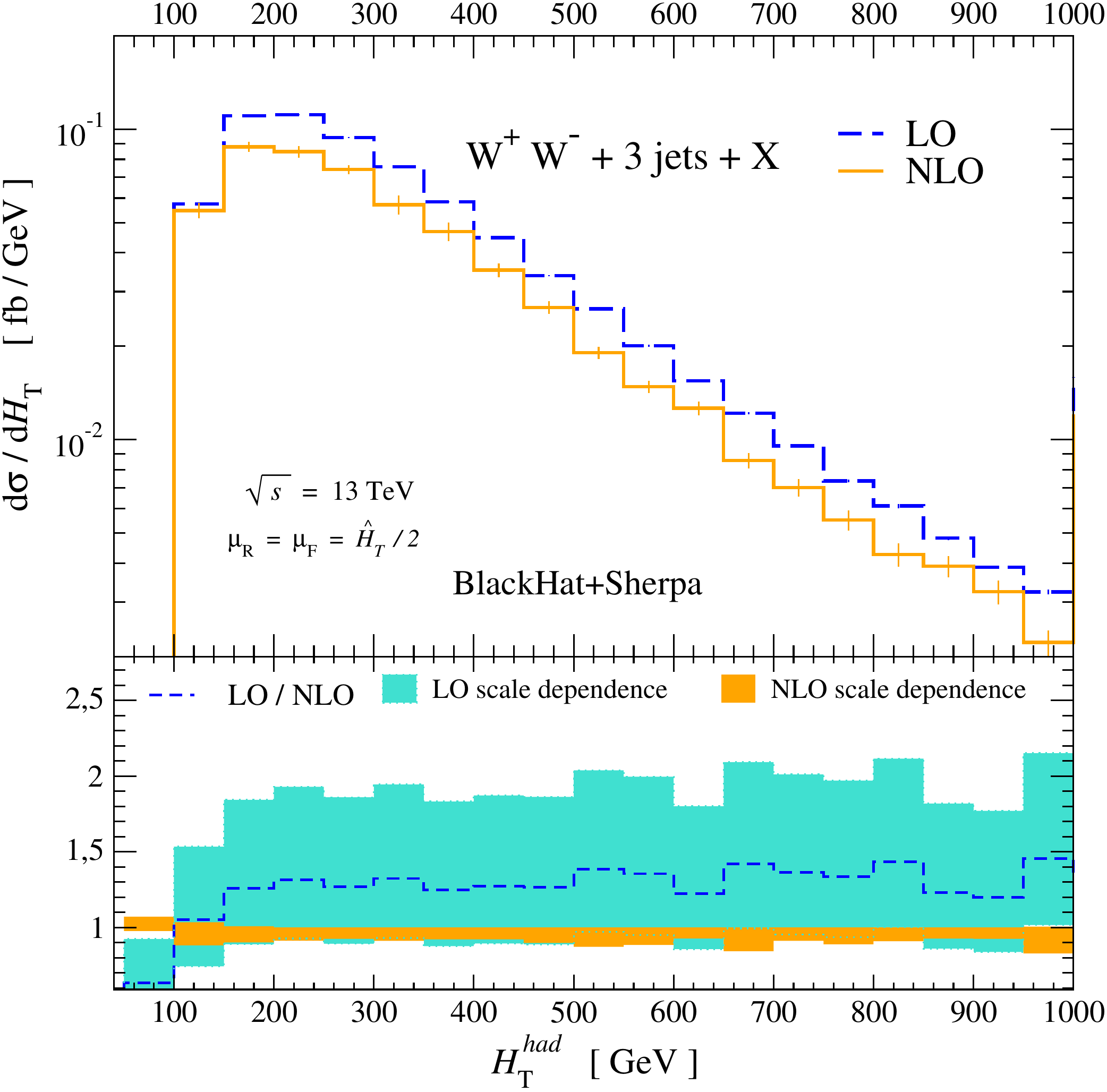}
\caption{The $H_{\rm T}^{had}$ distribution for \WpWmjjj{} jets at $\sqrt{s}=13$
TeV.
 }
\label{HTFigure}
\end{figure}

In \Fig{WW3ptFigure}, we show the $p_T$ spectra of the leading three jets in
\WpWmjjj-jet production at LO and NLO at $\sqrt{s}=13$ TeV; in the lower panels
we normalize (bin-by-bin) to the NLO prediction and show scale-dependence bands. 
We observe a
noticeable shape difference between the LO and NLO distributions for the first
two leading jets, while the shape for the third-jet distribution is very similar at LO and
NLO. This behavior has been observed in $W$ production in association with jets~\cite{W5jBH}.
\Fig{HTFigure} shows the distribution of the total hadronic transverse
energy $H_T^{had} = \sum_j E_{T,j}^{\rm jet}$.
We show the NLO and LO predictions, along with their scale-dependence bands. As
in the $p_T$ distributions, the NLO band is narrower.  The shapes of the
distributions at LO and NLO are similar, although they differ in their
normalization. 

With our new matrix elements for di-vector bosons production a number of
future directions for studies open up.  In order to compare the parton-level predictions to
experimental data, non-perturbative effects (such as hadronization and the
underlying event) have to be estimated. NLO parton-shower Monte Carlo programs
\cite{POWHEGBOXMCNLO} achieve this task. The virtual corrections computed here
can be incorporated into such studies.  Initially, cross section ratios like
the ones presented in \tab{CrossSectionLHC}, can provide useful predictions for
direct comparison to data, as parton-shower and non-perturbative effects are
largely canceled in them.
Furthermore, the results of this study show a good control over predictions to
\WpWmjjj-jet production within the SM. A number of immediate phenomenological
questions can be addressed.
It will be interesting, and necessary, to explore the effect of QCD corrections
for distinguished phase space regions requiring two forward tagging jets which
emphasize weak vector-boson-fusion event topologies as well as cuts for
new-physics searches.

The presented scattering process is but the first step to providing a new
class of predictions.  Beyond \WpWmjn{}-jet processes, additional combinations
of vector-boson pairs $VV^\prime$ (with $V$ and $V^\prime$ either $W^\pm$, $Z$
or a photon) as well as interference effects with signal processes play an
important role for exploring the electroweak symmetry breaking mechanism. These
processes are well within reach of the current \BlackHat{} library and we
expect to provide precise theory predictions in the near future.

\noindent {\bf Acknowledgements: } We thank Z.~Bern, L.J.~Dixon, S.~H\"oche,
D.A.~Kosower, D.~Maitre and C.~Schwinn for helpful discussions.  H.I.'s work is supported by a
Marie Sk{\l}odowska-Curie Action Career-Integration Grant PCIG12-GA-2012-334228
of the European Union.  P.H.'s work is supported by the Juniorprofessor Program
of Ministry of Science, Research and the Arts of the state of
Baden-W\"urttemberg, Germany.  The work of F.F.C. is supported by the Alexander
von Humboldt Foundation, in the framework of the Sofja Kovalevskaja Award 2014,
endowed by the German Federal Ministry of Education and Research.  This work
was performed on the bwUniCluster funded by the Ministry of Science, Research
and the Arts Baden-W\"urttemberg and the Universities of the State of
Baden-W\"urttemberg, Germany, within the framework program bwHP.

\end{document}